\documentclass[12pt,a4paper]{article}

\usepackage[T2A]{fontenc}
\usepackage[cp1251]{inputenc}
\usepackage[english]{babel}

\usepackage{amssymb,amsmath}
\usepackage{latexsym}
\usepackage{graphicx}
\usepackage{xcolor}
\usepackage{verbatim}

\begin{document}

\title{Electronic and thermoelectric properties of\\ quasi-fractal carbon nitride nanoribbons}


\author{R.T. Sibatov$^{1,2}$, Alireza Khalili Golmankhaneh$^3$, \\ R.M.~Meftakhutdinov$^4$, E.V.~Morozova$^4$, D.A.~Timkaeva$^4$}

\maketitle

\begin{center}
	$^{1}$ Moscow Institute of Physics and Technology (MIPT), Dolgoprudny 141700, Russia\\
	$^{2}$ Scientific-Manufacturing Complex ``Technological Centre'', Moscow 124498, Russia\\	
	$^3$ Department of Physics, Urmia Branch, Islamic Azad University,
	Urmia, Iran\\
$^4$ Ulyanovsk State University, Ulyanovsk 432017, Russia
\end{center}

\begin{abstract}
	{
		Recent works  devoted to the synthesis of artificial molecular systems with quasi-fractal geometry provide new opportunities for the experimental study of electronic properties in atomic systems of fractional dimension. 
There has been a renewed interest in theoretical studies of transport phenomena in fractal nanosystems.
In this work, using calculations based on the density functional theory, molecular dynamics, and the method of non-equilibrium Green functions, we study electronic and thermoelectric properties of a device based on quasi-fractal carbon nitride nanoribbon with Sierpinski triangle blocks. 
	We estimate changes in transport properties with generation $g$ of quasi-fractal blocks in nanoribbon. 
}
\end{abstract}

\section{Introduction}

The dimension of the electronic quantum system largely determines its properties. Particularly, in one-dimensional systems electrons form the Luttinger liquid~\cite{haldane1981luttinger}, and in two-dimensional systems the quantum Hall effect is observed~\cite{zhang2005experimental}. Little is known about the behavior of electrons in fractional dimensional systems \cite{kempkes2019}. 
Recent works \cite{li2017packing, kempkes2019, li2017construction} devoted to the synthesis of artificial molecular systems with quasi-fractal geometry provide new opportunities for experimental study of electronic properties in atomic systems of fractional dimension. The molecular quasi-fractal in the form of Sierpinski triangle was experimentally obtained in~\cite{li2017construction} by self-assembly of organic molecules by halogen bonds, hydrogen bonds, covalent bonds, and coordination interactions of metal-organic compounds on the surface. The resulting structure contains pores of different sizes in one triangular block and, according to forecasts, it has unique optical, magnetic and mechanical properties.

In works \cite{ghadiyali2019confinement, westerhout2018plasmon, pedersen2020graphene}, the first principles calculations and semi-empirical methods are used for the theoretical study of electronic properties of fractional-sized molecular systems.
In \cite{pedersen2020graphene},  electronic, optical, and magnetooptic properties of the Sierpinski fractal triangles built on a graphene sheet are studied within the Hubbard model.
Significant differences in the properties of structures with ``zigzag'' and ``armchair'' edges were found, similar to nanoribbons, quantum dots, and anti-dots. Sierpinski triangles with zigzag edges have a large proportion of boundary states, that leads to instability with respect to spin polarization  with account for the electron-electron interaction. On the contrary, fractals with armchair edges remain balanced on spin. In both cases, there is a strong energy gap, which leads to a pronounced optical absorption in the visible energy range. It has been shown that the distribution of energy levels is self-similar for quasi-fractal sets of later generation. 
In~\cite{khalili2021fractal}, the authors suggest a fractal Kronig-Penny model describing the quantum behavior of a particle in fractal lattice.
In~\cite{xu2021quantum}, the properties of photon transport in fractal grids in the form of Sierpinski carpet and triangle are experimentally examined,  the mean squared displacement  and the Polia number are determined. The authors observe anomalous diffusion of photons, and determine the dependence of the critical point of transition from normal to anomalous transfer on the fractal geometry parameters.

The authors of Ref.~\cite{li2017packing} implemented the packing of molecular Sierpinski triangles in one-dimensional crystals by the templates method in ultra-high vacuum. The obtained structures were studied by means of low-temperature scanning tunneling microscopy. The states described by electron wave functions of  fractional dimension were observed. The wave functions delocalized over the Sierpinski structure are spread out on self-similar parts at higher energies, and this large-scale invariance can also be recovered in the reciprocal space.

\begin{figure}[h]
	\centering
	\includegraphics[width=0.4\textwidth]{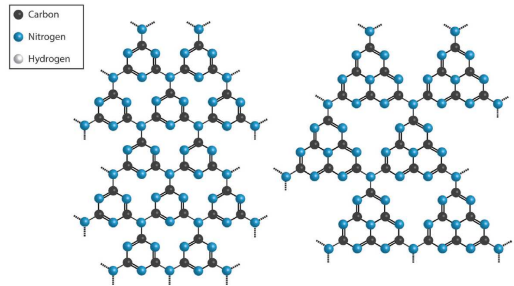}
	\caption{Triazine-based structure C$_3$N$_4$ (left panel) and poly-heptazine structure (tri-s-triazine) (right panel) \cite{miller2017carbon}.}
	\label{fig_CN}
\end{figure}

We consider nanoribbons constructed from quasi-fractal triangles. The geometry of these ribbons is similar to the geometry of the experimental structures demonstrated in~\cite{li2017packing}, but we have chosen carbon nitride structures as structural blocks.
Recently, nitrogen and carbon compounds with high N:C ratio and graphite polymer structure have been actively explored as potential next-generation materials for inclusion in energy conversion and storage devices, as well as for optoelectronic and catalytic applications \cite{miller2017carbon}.
Among them, popular materials consist of C- and N-containing heterocycles with heptazine or triazine rings bound via sp$_2$-linked nitrogen atoms N(C)$_3$ or --NH-- groups~\cite{miller2017carbon, bafekry2019two} (Fig.~\ref{fig_CN}).
In some works (see, e.g., \cite{bafekry2021two}), the authors consider structures where the central nitrogen atom in a heptazine triangle is substituted with a carbon atom.

\begin{figure}[tbh]
	\centering
	\includegraphics[width=0.95\textwidth]{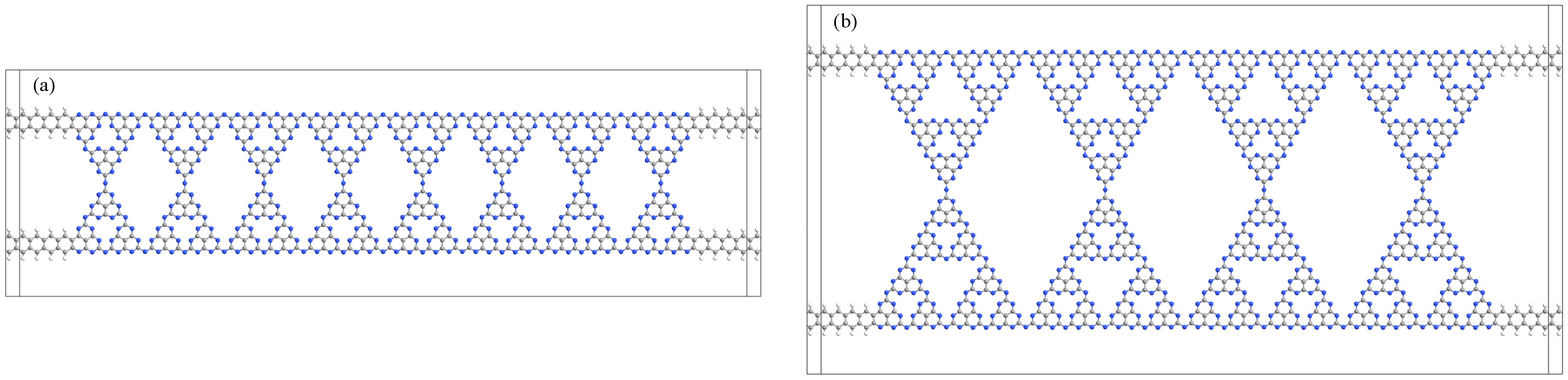}
	\caption{The geometry of devices based on carbon nitride nanoribbons with the block triangles of first and second generation.}
	\label{fig_geometry}
\end{figure}

In this paper, we study electron and phonon transport in a device based on a quasi-fractal carbon nitride nanoribbons with quasi-fractal Sierpinski triangles, using calculations based on methods of density functional theory (DFT), molecular dynamics (MD) and non-equilibrium Green functions (NEGF). The cases of the first and second generations of such ribbons are shown in Fig.~\ref{fig_geometry}. As electrodes, we used graphene nanoribbons attached to the corners of the studied quasi-fractal ribbons, as shown in Fig.~\ref{fig_geometry}. We calculate the electron and phonon transmission spectra, and then estimate the thermoelectric properties of the nanoribbons under consideration for different generations $g$.
In the second part of the work, using the Monte Carlo simulation method, we investigate the properties of hopping transport in quasi-fractal lattices, determine the mode and parameters of anomalous diffusion in the absence and presence of an external electric field for different values of localization radius and levels of energy disorder.

\section{Electronic properties}

Let us first consider the flat quasi-fractal crystalline lattices by performing their optimization and calculating their electronic band structures and absorption spectra. These lattices are periodic with an elementary cell, representing the atomic quasi-fractal Sierpinski triangle of $g$-th generation. Optimization is performed for quasi-fractal crystalline structures using the DFT method implemented in Quantum ATK~\cite{smidstrup2019quantumatk}. The cutoff energy of the electron wave functions is 500 eV. The reliable convergence criteria for total energy and power are as follows $10^{-6}$ eV and 0.01 eV/\AA. The transverse cell size is 30~\AA. We used the Perdue-Burke-Ernzerhoff approximation to describe the effects of electronic exchange and correlation, to optimize the structure and to calculate electronic properties. The Monkhorst--Pack method was used to generate $k$ points in the Brillouin zone.

\begin{figure}[h]
	\centering
	\includegraphics[width=0.85\textwidth]{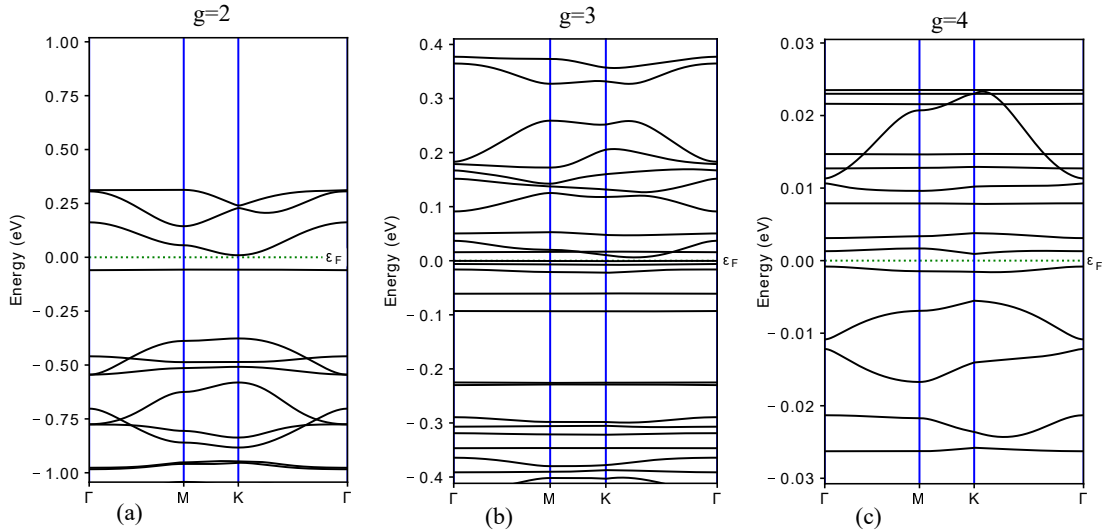}
	\caption{Band structures of quasi-fractal lattices for $g=2$ (a), $g=3$ (b), $g=4$ (c).}
	\label{fig_bands}
\end{figure}

\begin{figure}[h]
	\centering
	\includegraphics[width=0.65\textwidth]{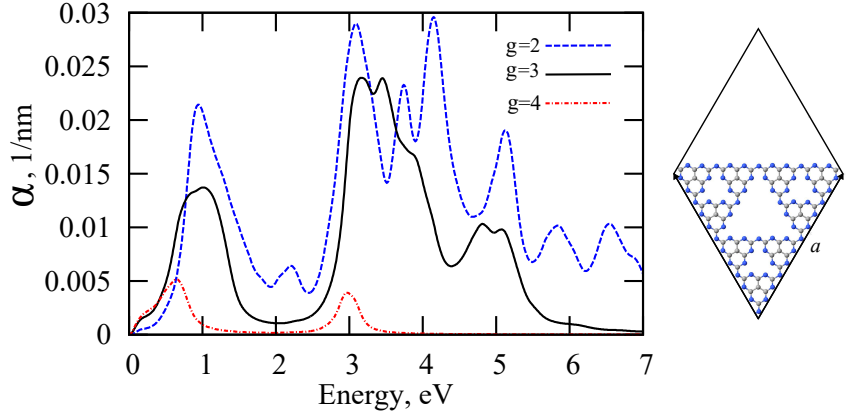}
	\caption{Absorption spectra of planar quasi-fractal crystals for three generations. The right panel shows the structure for the case $g = 3$; the crystal contains triangular pores of three sizes.}
	\label{fig_absorption}
\end{figure}

Fig.~\ref{fig_bands} shows the band structures of the periodic nitrogen-carbon lattices, in which the unit cell contains a quasifractal Sierpinski triangle  of $g$-th generation. Fig.~\ref{fig_absorption} shows the absorption spectra of the corresponding structures. To the right of Fig.~\ref{fig_absorption}, a unit cell for case $g = 3$ is shown. In the crystal with $g=3$, there are triangular pores of three sizes. With the increase of $g$ for the elementary blocks of the crystal structure, the flattening of the branches is observed and, consequently, the increase in the effective masses of the charge carriers takes place. Individual zones for large $g$ become simple lines of the molecular system  due to the localization of the carriers in fragments of the quasi-fractal blocks.

Usually fractal systems are characterized by a wider absorption frequency range. In some sense, our results contradict this view. The absorption spectra in Fig.~\ref{fig_absorption} indicate that with an increase in $g$ not only the absorption index decreases. The absorption bands are constricted and the corresponding structures are characterized by a narrower absorption range. This is due to the fact that, in comparison with macroscopic fractals, in atomistic structures, along with a change in geometry, the electronic states of the system change. The spectrum of the quasi-fractal system under consideration tends to go from the spectrum of the periodic 2d system to the spectrum of an isolated molecular system.

\begin{figure}[h]
	\centering
	\includegraphics[width=0.85\textwidth]{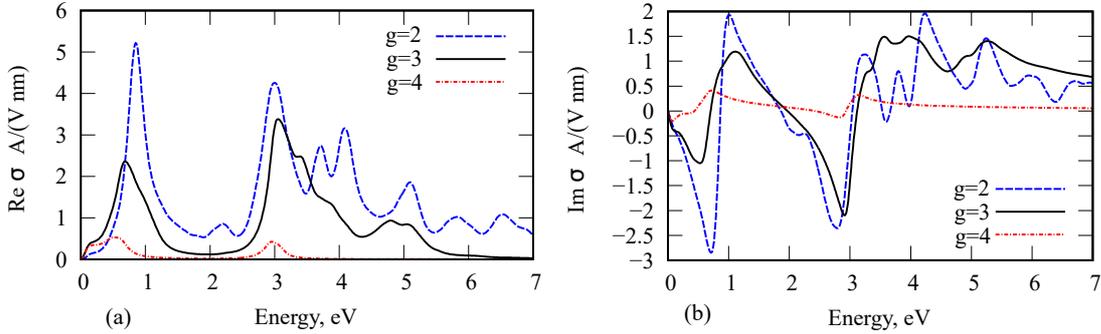}
	\caption{Real and imaginary parts of the optical conductivity of quasi-fractal gratings for $g=2$ (a), $g=3$ (b), $g=4$ (c).}
	\label{fig_sigma}
\end{figure} 

Fig.~\ref{fig_sigma} shows the frequency dependencies of the real and imaginary parts of the optical conductivity for the investigated quasi-fractal crystal structures.
As the generation $g$ increases, the optical conductivity decreases and the nonzero conduction bands become narrower. The imaginary part of conductivity also decreases for larger $g$. 

\begin{figure}[h]
	\centering
	\includegraphics[width=0.95\textwidth]{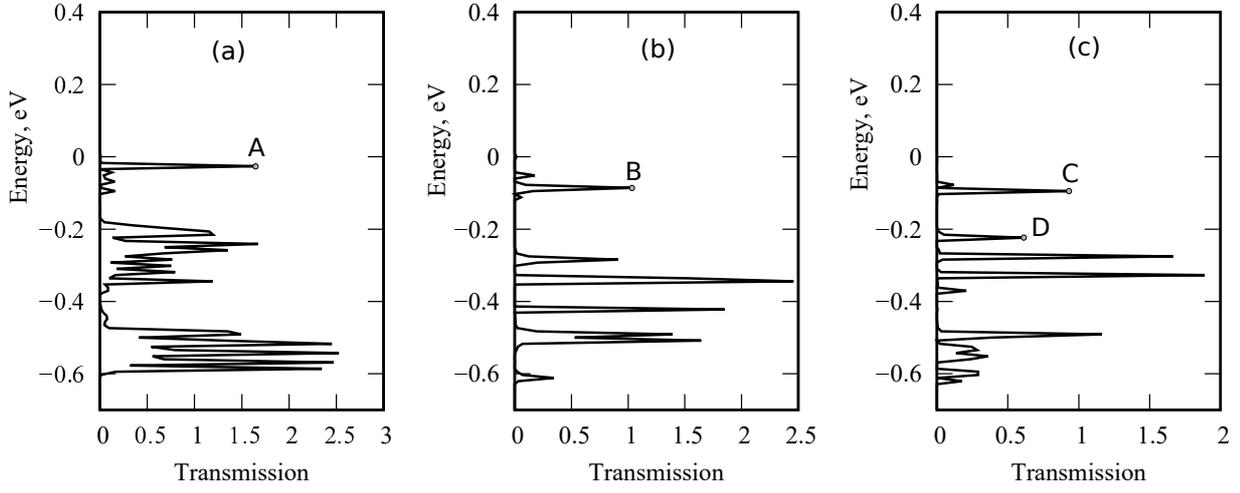}
	\caption{Transmission spectra of devices based on a nitrogen-carbon quasi-fractal nanoribbon for generations $g=1$ (a), 2 (b), and 3 (c).}
	\label{fig_transmission}
\end{figure}

\begin{figure}[h]
	\centering
	\includegraphics[width=1\textwidth]{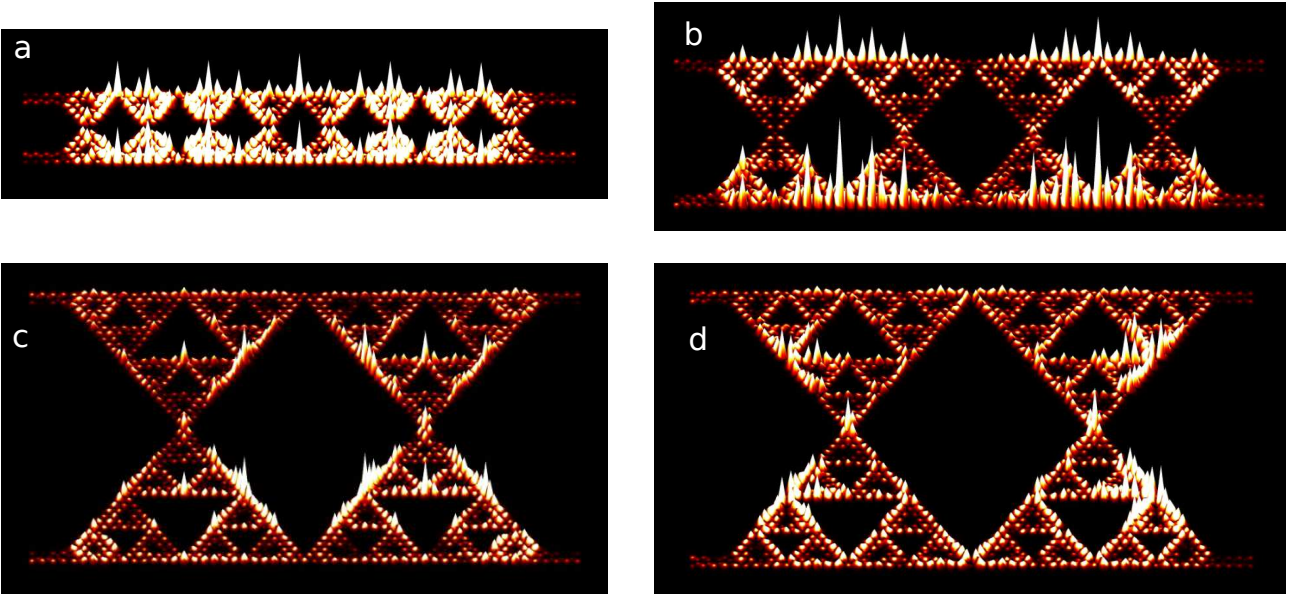}
	\caption{Transfer eigenstates in quasi-fractal nitrogen-carbon nanoribbons. The states correspond to points A, B, C, and D on the transmission spectra of the quasi-fractal nanoribbon for generations $g=1$ (a), 2 (b), and 3 (c,d).}
	\label{fig_eigenstates}
\end{figure} 

In quasi-fractal nanoribbons, we also observe a transition from transmission bands to narrow transmission channels (Fig. \ref{fig_transmission}).
When calculating the transmission spectra, nanoribbons were chosen with the same length equal to $L = 109.5$~\AA, but with a different number of quasi-fractal units. In QuantumATK, the transfer matrix is calculated according to the following formula
$$T_{nm}(E,\mathbf{k}) = \sum_\ell t_{n\ell}(E,\mathbf{k})  t^\dagger_{\ell m}(E,\mathbf{k}),$$
where $t_{nk}$ is the amplitude of the transfer from the Bloch state $\psi_n$ in the left electrode to the Bloch state $\psi_k$ in the right electrode, the matrix $t^\dagger$ is an Hermitian conjugate. The transmittance is defined as the trace of the transmission matrix,
$$T(E,\mathbf{k}) = \sum_n T_{nn}(E,\mathbf{k}).$$
Let $\lambda_\alpha$ be the eigenvalues of the transfer matrix $T_{nm}$.
From the invariance of the trace of the matrix:
$$T(E,\mathbf{k}) = \sum_\alpha \lambda_\alpha(E,\mathbf{k}),$$
where $\lambda_\alpha\in [0,1]$ are transmission eigenvalues for each spin channel.

The transmission eigenstates are calculated by diagonalizing a linear combination of Bloch states, $\sum_n e_{\alpha,n} \psi_n$, where $e_{\alpha,n}$
are vectors of the basis diagonalizing the transmission matrix: 
$$\sum_m T_{nm}e_{\alpha,m}= \lambda_\alpha e_{\alpha,n}.$$

Fig.~\ref{fig_transmission} shows the transmission spectra for devices based on a nitrogen-carbon quasi-fractal nanoribbon for generations $g = 1$, 2 and 3.
Fig.~\ref{fig_eigenstates} demonstrates examples of transfer eigenstates in quasi-fractal nitrogen-carbon nanoribbons. These states correspond to the peaks in the transmission spectra marked with points A, B, C and D in Fig.~\ref{fig_transmission}. The transmission eigenstates are distributed throughout the structure and take on maximum values at the edges of the quasi-fractal structure.

\section{Thermoelectric properties}

In this section, we estimate thermoelectric properties of quasi-fractal nanoribbons.
The search for new thermoelectric nanosystems is an important area of research associated with possible applications in power generation and cooling systems on nanoscales. Several studies point to the important role of quasi-one-dimensional geometry in improving the thermoelectric properties of \cite{ouyang2009theoretical, kodama2017modulation} nanosystems. 

The maximum efficiency of the energy conversion process in a thermoelectric material is determined by its thermoelectric figure of merit $ZT$, determined by the expression
$$
ZT=\frac{S^{2}GT}{\lambda},
$$
where $S$ is the Seebeck coefficient, $G$  electrical conductivity, $T$ absolute temperature, $\lambda$ thermal conductivity, which is equal to the sum of the electron $\lambda_{e}$ and phonon $\lambda_\mathrm{ph}$ thermal conductivity.

The transfer coefficients were calculated using the NEGF-method, DFT, and nonequilibrium molecular dynamics. We used a model in which the central part (quasi-fractal nanoribbon) is connected to semi-infinite left and right electrodes (graphene nanoribbons). QuantumATK \cite{smidstrup2019quantumatk} calculates the specified thermoelectric coefficients and Peltier coefficient according to the linear response theory. The following relationships are used
$${G_{\rm{e}}} = {\left. {{{dI} \over {d{V_{{\rm{bias}}}}}}} \right|_{dT = 0}},\ S =  - {\left. {{{d{V_{{\rm{bias}}}}} \over {dT}}} \right|_{I = 0}},\ {\lambda _{\rm{e}}} = {\left. {{{d{I_Q}} \over {dT}}} \right|_{I = 0}},\ \Pi  = {\left. {{{{I_Q}} \over I}} \right|_{dT = 0}} = S{V_{{\rm{bias}}}}.$$
Here $I_Q = dQ / dT$ is the electronic component of the heat flux.

We use DFT to calculate electron transmission and MD method to calculate phonon transmission. In the MD method, we use the optimized empirical potential ReaxFF (2015) \cite{pai2016development} for the molecular dynamics of systems containing carbon C, nitrogen N, hydrogen H, and boron B (the latter component is absent in our case).
Previously, similar computational tools have been successfully applied to a number of carbon and non-carbon nanoscale systems (see e.g.~\cite{markussen2009electron} for details).

\begin{figure}[h]
	\centering
	\includegraphics[width=0.95\textwidth]{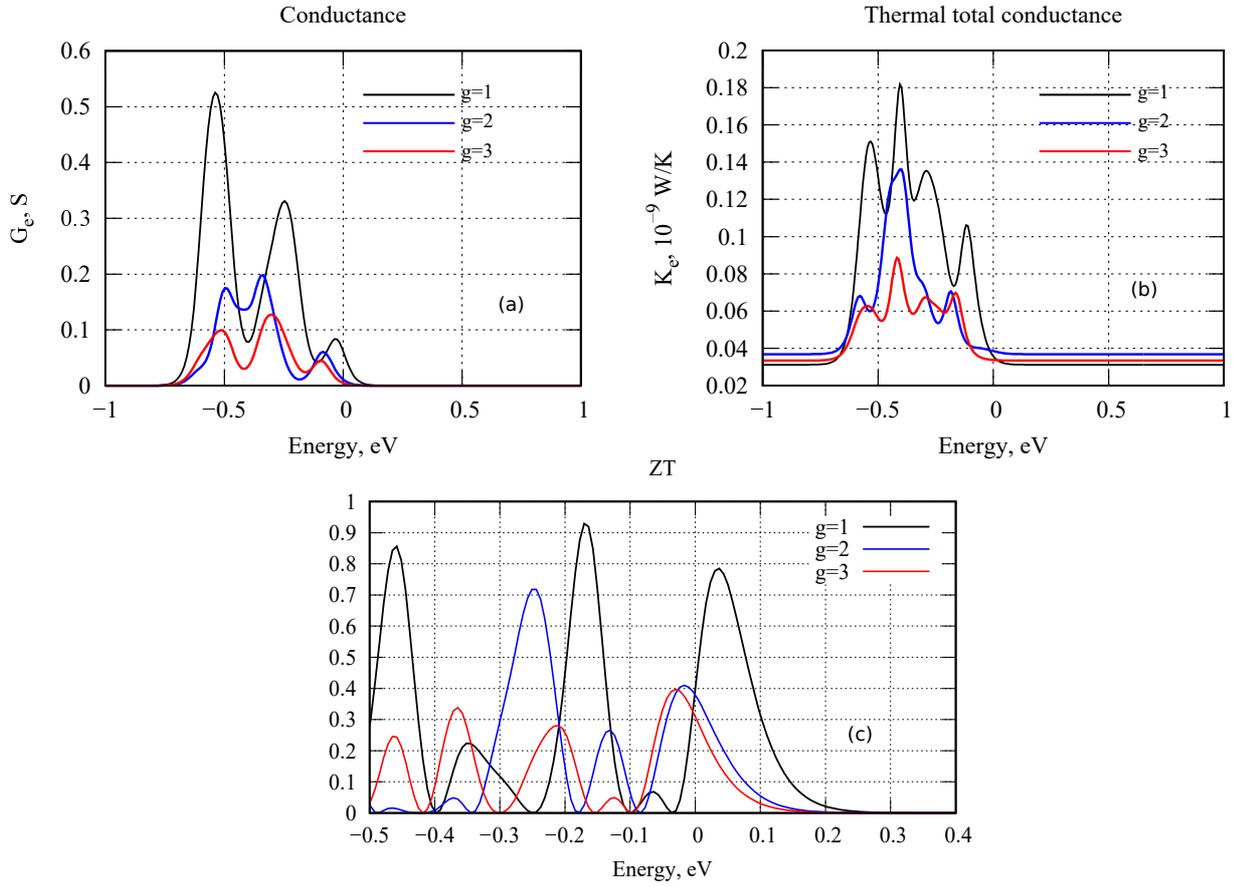}
	\caption{Conductance (a), thermal conductivity (b) and thermoelectric figure of merit (c) of devices based on quasi-fractal nitrogen-carbon nanoribbons.}
	\label{fig_thermo}
\end{figure}

Fig.~\ref{fig_thermo} shows conductance, thermal conductivity and thermoelectric figure of merit of devices based on quasi-fractal nitrogen-carbon nanoribbons with molecular quasi-fractal blocks of three generations. As can be seen from the electronic transmission spectra, with an increase in the number of generations $g$, the conductivity decreases due to the localization of electrons in fragments of quasi-fractal system.
Thermal conductivity due to phonons does not behave monotonically with a change
of $g$: $\lambda_\mathrm{ph}(E;g=1)<\lambda_\mathrm{ph}(E;g=3)<\lambda_\mathrm{ph}(E;g=2)$. 
The obtained values of the thermoelectric figure of merit are several tenths, and at some values of the energy are close to 1, which indicates the possible use of the considered nanoribbons for thermoelectric applications, although it should be noted that with an increase in $g$ the characteristic values of ZT decrease.

\section{Conclusion}

The method of packaging molecular Sierpinski  triangles in one-dimensional crystals by the method of templates in a super high vacuum allows to create nanoribbons with quasi-fractal geometry. In this work, using calculations based on DFT, MD and NEGF  methods, we study electronic and phonon transport in a device based on a quasi-fractal carbon nitride nanoribbon with Sierpinski triangle blocks.

Usually fractal systems are characterized by a wider absorption frequency range. Contrary to this, we observe that the absorption bands are constricted for larger $g$ and the corresponding structures are characterized by a narrower absorption range. This is due to the fact that, in comparison with macroscopic fractals, in atomistic structures, along with a change in geometry, the electronic states of the system are modified. The spectrum of the quasi-fractal system under consideration tends to go from the spectrum of the periodic 2d system to the spectrum of an isolated molecular system. 

The obtained values of the thermoelectric figure of merit are several tenths, and at some values of the energy are close to 1, which indicates the possible use of the considered nanoribbons for thermoelectric applications, although it should be noted that with an increase in $g$ the characteristic values of ZT decrease.

\section*{Acknowledgement}

The work is partially supported by the Ministry of Science and Higher Education of the Russian Federation (project 0591-2021-0002).

\end{document}